\documentclass[12pt,letterpaper]{article}
\usepackage{graphicx}
\usepackage{amssymb}
\usepackage{epstopdf}

\usepackage{amsmath,amssymb,array,calc,rotating,epsfig,psfrag}
\usepackage[nosort]{cite}

\numberwithin{equation}{section}

\def\be{\begin{equation}}
\def\ee{\end{equation}}
\def\bea{\begin{eqnarray}}
\def\eea{\end{eqnarray}}

\def\a{\alpha}
\def\b{\beta}
\def\c{\chi}
\def\d{\delta}
\def\e{\epsilon}

\def\g{\gamma}
\def\h{\eta}

\def\j{\psi}

\def\l{\lambda}

\def\o{\omega}

\def\q{\theta}
\def\r{\rho}
\def\s{\sigma}
\def\t{\tau}

\def\x{\xi}

\def\F{\Phi}
\def\G{\Gamma}

\def\L{\Lambda}
\def\O{\Omega}
\def\P{\Pi}

\def\del{\partial}

\def\mcal{\mathcal}

\def\ra{\rightarrow}

\def\pp{{\mathchoice
          {
              \kern 1pt%
              \raise 1pt
              \vbox{\hrule width5pt height0.4pt depth0pt
                    \kern -2pt
                    \hbox{\kern 2.3pt
                          \vrule width0.4pt height6pt depth0pt
                          }
                    \kern -2pt
                    \hrule width5pt height0.4pt depth0pt}%
                    \kern 1pt
           }
            {
              \kern 1pt%
              \raise 1pt
              \vbox{\hrule width4.3pt height0.4pt depth0pt
                    \kern -1.8pt
                    \hbox{\kern 1.95pt
                          \vrule width0.4pt height5.4pt depth0pt
                          }
                    \kern -1.8pt
                    \hrule width4.3pt height0.4pt depth0pt}%
                    \kern 1pt
            }
            {
              \kern 0.5pt%
              \raise 1pt
              \vbox{\hrule width4.0pt height0.3pt depth0pt
                    \kern -1.9pt  
                    \hbox{\kern 1.85pt
 \vrule width0.3pt height5.7pt depth0pt
                          }
                    \kern -1.9pt
                    \hrule width4.0pt height0.3pt depth0pt}%
                    \kern 0.5pt
            }
           {
              \kern 0.5pt%
              \raise 1pt
              \vbox{\hrule width3.6pt height0.3pt depth0pt
                    \kern -1.5pt
                    \hbox{\kern 1.65pt
                          \vrule width0.3pt height4.5pt depth0pt
                          }
                    \kern -1.5pt
                    \hrule width3.6pt height0.3pt depth0pt}%
                    \kern 0.5pt
            }
        }}

 \def\mm{{\mathchoice
                       {
                             \kern 1pt
               \raise 1pt    \vbox{\hrule width5pt height0.4pt depth0pt
                                  \kern 2pt
                                  \hrule width5pt height0.4pt depth0pt}
                             \kern 1pt}
                       {
                            \kern 1pt
               \raise 1pt \vbox{\hrule width4.3pt height0.4pt depth0pt
                                  \kern 1.8pt
                                  \hrule width4.3pt height0.4pt depth0pt}
                             \kern 1pt}
                       {
                            \kern 0.5pt
               \raise 1pt
                            \vbox{\hrule width4.0pt height0.3pt depth0pt
                                  \kern 1.9pt
                                  \hrule width4.0pt height0.3pt depth0pt}
                            \kern 1pt}
                       {
                           \kern 0.5pt
             \raise 1pt  \vbox{\hrule width3.6pt height0.3pt depth0pt
                                  \kern 1.5pt
                                  \hrule width3.6pt height0.3pt depth0pt}
                           \kern 0.5pt}
}}
 \def\ad{{\kern0.5pt
                  \alpha \kern-5.05pt \raise5.8pt\hbox{$\textstyle.$}\kern
 0.5pt}}
 \def\bd{{\kern0.5pt
                   \beta \kern-5.05pt \raise5.8pt\hbox{$\textstyle.$}\kern
 0.5pt}}

 \def\qd{{\kern0.5pt
                   q \kern-5.05pt \raise5.8pt\hbox{$\textstyle.$}\kern
 0.5pt}}
 \def\Dot#1{{\kern0.5pt
     {#1} \kern-5.05pt \raise5.8pt\hbox{$\textstyle.$}\kern
 0.5pt}}

\def\md{{\mathchoice
   {
      {{\kern 1pt - \kern-6.2pt \raise5pt\hbox{$\textstyle.$}\kern 1pt}}}
    {
      {{\kern 1pt - \kern-6.2pt \raise5pt\hbox{$\textstyle.$}\kern 1pt}}}
    {
      {\kern0.5pt - \kern-5.05pt \raise3.4pt\hbox{$\textstyle.$}\kern0.5pt}}
    {
      {\kern0.5pt - \kern-5.05pt \raise3.4pt\hbox{$\textstyle.$}\kern0.5pt}}}}

\def\pd{{\kern0.5pt
                   + \kern-5.05pt \raise5.8pt\hbox{$\textstyle.$}\kern
0.5pt}}

\def\pmd{{\kern0.5pt
                  \pm \kern-5.05pt \raise6.3pt\hbox{$\textstyle.$}\kern1.5pt}}

\begin{document}
\begin{titlepage}
\rightline{\tt UMDEPP-07-002}
\begin{center}
{\Large \bf  D = $2$, $\mcal{N}=(2,2)$}\\
[0.1in]
{\Large \bf Semi Chiral Vector Multiplet}\\
[1.25in]
{\Large S. James Gates, Jr.,~~~Willie Merrell}\\
[0.25in]
{\it Center for String and Particle Physics}\\
[0.05in]
{\it Department of Physics, University of Maryland}\\
[0.05in]
{\it College Park, MD 20742}\\
\begin{abstract}
We describe a new $1+1$ dimensional $\mcal{N}=(2,2)$ vector multiplet that naturally couples to semi chiral superfields in the sense that the gauged supercovariant derivative algebra is only consistent with imposing covariantly semi chiral superfield constraints.  It has the advantages that its prepotentials shift by semi chiral superfields under gauge transformations.  We also see that the multiplet relates the chiral vector multiplet with the twisted chiral vector multiplet by reducing to either multiplet under appropriate limits without being reducible in terms of the chiral and twisted chiral vector multiplet.  This is explained from the superspace geometrical point of view as the result of possessing a symmetry under the discrete supercoordinate transformation that is responsible for mirror copies of supermultiplets.  We then describe how to gauge a non linear sigma model with semi chiral superfields using the prepotentials of the new multiplet.
\end{abstract}
\end{center}
\end{titlepage}
\newpage
\section{Introduction}
The representations of 2d $\mcal{N}=(2,2)$ superfields play an important role in understanding string theory, complex geometry, and the interface between the two subjects.  The seminal work of Zumino \cite {zumino} showed the connection between $\mcal{N}=(2,2)$ non linear sigma models (NLSM), with chiral superfields and K\"{a}hler geometry.  Gates, Hull, and Ro\v{c}ek \cite{ghr} formally introduced the twisted chiral superfield representation that was implicitly containted in \cite{gates2} and discovered a NLSM that realized a bi-hermitian geometry with commuting complex structures.  The bi-hermitian geometries contain NS-NS two form flux and are examples of generalized K\"{a}hler geometries \cite{hit, thesis}.  A sigma model realization of bi-hermitian geometry with non commuting complex structures was obtained by the introduction of left and right semi chiral superfields \cite{blr}.  These sigma models where also shown to be generalized K\"{a}hler \cite{lruz}.  Later it was shown in \cite{lruz2} that all generalized K\"{a}hler geometries can be described locally by a $\mcal{N}=(2,2)$ sigma model with chiral, twisted chiral, left and right semi chiral superfield representations.  

The vector multiplets also play an important role in the connection between string theory and complex geometry.  Gauging the isometries of a NLSM is a necessary part of understanding T duality and requires the use of certain target space background data.  If the target space has NS-NS flux $H=db$, then a one form,$u$, arises from the conditon that the isometry preserves the NS-NS flux, $\mcal{L}_\x H=0$, which implies that $\mbox{i}_\x H=du$.  The one form $u$ is needed to gauge the sigma model \cite{hs}.  Here $\x$ is the killing vector for the isometry.  In order to gauge a sigma model with greater than $(1,0)$ or $(0,1)$ supersymmetry, one must use the moment map associated to the isometry \cite{bw,hklr,hps}.  There are two known vector multiplets for $(2,2)$ supersymmetry, the chiral and the twisted chiral vector multiplet.  Here we identifiy the vector multiplet by the transformations of the prepotentials, i.e. the chiral multiplet has a prepotential that shifts by a chiral superfield under gauge transformations and like wise for the twsited chiral multiplet.  The known vector multiplets explain the dualtiy between chiral and twisted chiral superfields and provide the manifestly $(2,2)$ description of T duality \cite{rv} which is the essence of mirror symmetry \cite{syz}.  However the complete story of T duality isn't know for the most general $(2,2)$ NLSM.  In \cite{mpv,kt} the chiral vector multiplet is coupled to the semi chiral representations at the manifest $(2,2)$ level and extended $(1,1)$ level on shell.  In \cite{mpv}, the discussion of T duality requires the introduction of extra fields that result in a degenerate metric.  Starting with a real dimension 2n space, one is forced to think of the T dual geometry as a codimension 2 surface of a 2n+2 space.  In a similar way, one would encounter problems with quotients of target spaces described by semi chiral superfields.  The portion of a sigma model parameterized by semi chiral superfields is necessarily of $4n$ real dimensions with $n$ some integer \cite{st}.  However constructing a quotient using the chiral or twisted chiral vector multiplet would result in a sigma model of $4n-2$ real dimesions.  Such a quotient can't be described using semi chiral superfields and it is unclear if the remaining degree's of freedom can be assembled into chiral or twisted chiral representations.  This hints at a lack of general understanding of how to gauge a sigma model with semi chiral representations participating in the isometry.  The goal of the work is explore the issue of gauging sigma models with semi chiral superfield representations as a first step in developing a better description of T duality and the construction of quotients.  The approach taken is to ask if there is a vector multiplet that has same relationship to semi chiral superfields that chiral and twisted chiral vector multiplets have to chiral and twisted chiral matter.  Specifically, is there a vector multiplet with prepotentials that gauge transform by semi chiral superfields?  This question can also be asked covariantly as, is there a gauged supercovariant derivative algebra that is only compatible with defining covariantly semi chiral superfields?

From the perspective of superspace representation theory the work presented can be viewed in a different way.  The vector multiplet that we derive gives an example of a multiplet that possess a symmetry under the discrete transformation on superspace coordinates that is responsible for the existence of `mirror copies' of 2d $\mcal{N}=(2,2)$ supermultiplets.

The paper is organized as follows.  In section 2 we give the $U(1)$ gauged supercovariant derivative algebra and Bianchi identities for the semi chiral vector multiplet and discuss the gauge transformations of the prepotentials.  We also discuss the origin of the duality between the chiral and twisted chiral vector multiplets described in \cite{gates}.  In section 3 we present the discussion of the discrete transformation on superspace and the relevance of the semi chiral vector multiplet.   In section 4 we give the kinetic terms in the action for the semi chiral vector multiplet.  In section 5, we use the prepotentials to gauge a NLSM with $U(1)$ symmetry.  In section 6 we give summary and conclusion.

\section{The Semi Chiral Vector Multiplet}
The known irreducible scalar superfield representations in 2d $\mcal{N}=(2,2)$ are chiral, twisted chiral, and left/right semi chiral.  These superfields are defined using the supercovariant derivatives $D_A=(D_\a,\bar D_\a, \del_a)$ with algebra
\be
[D_\a,D_\b\}=0,~~~~[D_\a,\bar D_\b\}=2i(\g^a)_{\a\b}\del_a~~.
\ee
The representations are distinguished by the following supercovariant derivative constraints.
\bea
&&\mbox{Chiral:  } \bar D_\a \phi = 0\cr
\cr
&&\mbox{Twisted Chiral:  } \bar D_- \c = D_+ \c = 0\cr
\cr
&&\mbox{Left Semi Chiral:  } \bar D_+ X = 0\cr
\cr
&&\mbox{Right Semi Chiral:  } \bar D_- Y = 0
\eea
The semi chiral constraints are novel in the sense that they are only compatible with Lorentz invariance in 2 dimensions.  The vector multiplets that couple to chiral and twisted chiral matter are known and have been discussed \cite{hps,gates}.  One can observe that both the chiral and twisted chiral vector multiplets can couple to left/right semi chiral matter.  A natural question to ask is whether there is a vector multiplet for left/right semi chiral matter other than the chiral and twisted chiral vector multiplets.  To pursue this question we search for a gauge supercovariant derivative algebra that is only consistent with setting covariant left/right semi chiral constraints.
\subsection{The Algebra and B.I.'s}
We start by introducing gauge supercovariant derivatives $\nabla_A = D_A -i \G_A t$, where $D_A$ is given above, $\G_A$ is the supergauge field and $t$ is the abstract generator of the $U(1)$ symmetry we wish to gauge.  We then impose the following constraints on the gauge supercovariant derivative algebra.  For conventional constraints we impose the condition
\be
(\g_a)^{\a\b}[\nabla_\a,\bar\nabla_\b\}=-4i\nabla_a
\ee
The constraints that preserve semi chiral representations are
\be
\label{rep}
(\g_a)^{\a\b}[\nabla_\a,\nabla_\b\}=0.
\ee
The algebra and bianchi identites for the above constraints are
\bea
\label{algebra}
&&[\nabla_\a,\nabla_\b\}=4ig(\g^3)_{\a\b}\bar T t\cr
&&[\nabla_\a,\bar\nabla_\b\}=2i(\g^c)_{\a\b}\nabla_c+2g[C_{\a\b}S-i(\g^3)_{\a\b}P]t\cr
&&[\nabla_\a,\nabla_b\}=g(\g_b)_\a{}^\b \bar W_\b t-g(\g^3\g_b)_\a{}^\b\bar\O_\b t\cr
&&[\nabla_a,\nabla_b\}=-ig\epsilon_{ab}\mcal{W}t
\eea
and
\bea
\label{BI}
\nabla_\a S&=&-i\bar W_\a\cr
\cr
\nabla_\a P&=&-(\g^3)_\a{}^\b\bar W_\b\cr
\cr
\bar\nabla_\a T &=& 0\cr
\cr
\nabla_\a T &=&\O_\a \cr
\cr
\nabla_\a \O_\b &=& -C_{\a\b} \s \cr
\cr
\nabla_\a \bar\O_\b &=& 2i(\g^a)_{\a\b}\nabla_a \bar T \cr
\cr
\nabla_\a \bar W_\b&=&0\cr
\cr
\nabla_\a W_\b &=& iC_{\a\b}d-(\g^3)_{\a\b}(\s_1+\mathcal{W})+(\g^a)_{\a\b}\nabla_a S-i(\g^3\g^a)_{\a\b}\nabla_a P\cr
\cr
\nabla_\a d&=& (\g^a)_\a{}^\b\nabla_a \bar W_\b\cr
\cr
\nabla_\a \s &=& 0 \cr
\cr
\bar \nabla_\a \s &=& 2i(\g^a)_\a{}^\b\nabla_a\O_\b
\eea
where $\s_1=Re(\s)$.  It is of interest to note that the B.I.s require that $T$ is chiral and $\P=S-iP$ is twisted chiral.  At first glance one might think that this algebra is the direct sum of the algebra's for a chiral and twisted chiral vector multiplet.  That this isn't the case can be seen in at least two different ways.  The first is mixing of the auxiliarly field $\s$ in the B.I.'s, specifically in the $\nabla_\a W_\b$ and $\nabla_\a \O_\b$ terms.  The discussion of the second argument is better suited to take place after the discussion of prepotentials.
\subsection{Prepotentials}
The description given above is an off shell description and thus the field strengths can be solved for in terms of unconstrained prepotentials.  To find the prepotentials we consider the representation preserving constraint (\ref{rep}) and see what they imply for the potentials $\G_\a$.  In terms of the super field strengths we have
\bea
F_{++}&=&2D_+\G_+=0\ra \G_+=D_+\bar V_1\cr
\cr
F_{--}&=&2D_-\G_-=0\ra \G_-=D_-\bar V_2
\eea
The complex scalar fields $V_1$ and $V_2$ are the prepotentials.  The prepotentials have two types of gauge transformation.  Since the super field strengths are invariant under $\G_A\ra\G_A+D_A L$ where $L$ is an arbitrary real superfield, this implies that $V_1$ and $V_2$ share a common gauge transformation 
\be
V_1\ra V_1+L, ~~~~V_2\ra V_2+L
\ee
$V_1$ and $V_2$ also have a priori independent gauge transformations.  For $\bar D_+\L=0$ and $\bar D_-U=0$ the super field strengths are invariant under the transformations
\be
\label{prepot1}
V_1\ra V_1+\L
\ee
and
\be
\label{prepot2}
V_2\ra V_2+U
\ee
Here we see that we have found a vector multiplet with prepotentials that shift by semi chiral superfields under gauge transformations.  At this point we can also give the second argument as to why the semi chiral vector multiplet can't be obtained as a direct sum of the chiral and twisted chiral vector multiplets.  Recall for the chiral vector multiplet that after fixing the gauge symmetry parameterized by the analog of the $L$ gauge transformation, it has only one real prepotential.  The same is true for the twisted chiral vector multiplet.  One would expect that a direct sum of the chiral and twisted chiral vector multiplet would be described in terms of two real prepotentials.  However, for the semi chiral vector multiplet given above, there are three real prepotentials after $L$  gauge fixing.

The field strengths are given in terms of the prepotentials as
\bea
T&=&\frac{1}{4}\bar D^2(V_2-V_1)\cr
\cr
\bar T&=&\frac{1}{4}D^2(\bar V_2-\bar V_1)\cr
\cr
\P&=&S-iP=\frac{1}{2}D_+\bar D_-(V_2-\bar V_1)\cr
\cr
\bar\P&=&S+iP=\frac{1}{2}D_-\bar D_+(\bar V_2-V_1)
\eea
\subsection{Duality between Chiral and Twisted Chiral Vector multiplets}
While the semi chiral vector multiplet isn't reducible in terms of the chiral and twisted chiral vector multiplet, it contains both the chiral and twisted chiral vector multiplet.  This can be seen in the following way.  Starting with equation (\ref{algebra}) and setting the field strength $\bar T=0$, one finds that the B.I.'s require that $\O_\a=\s=0$.  The resulting algebra and B.I.'s are identical to the those for the chiral vector multiplet \cite{gates}.  Similarly if one sets $S=P=0$ then the B.I.'s require that $W_\a=d=0$ and $\s_1=-\mcal{W}$.  This then gives the algebra and B.I.'s for the twisted chiral vector multiplet.  In this way we can view the semi chiral vector multiplet as the parent multiplet that gives rise to the chiral and twisted chiral vector multiplet.  This isn't very surprising in hind sight.  The semi chiral constraint is weaker that the chiral or twisted chiral constraint.  It is only the zero modes allowed for a massless representation that distinguishes a semi chiral superfield from the sum of a chiral and twisted chiral superfield.  From this point of view, one could expect the semi chiral vector multiplet to incorporate both the chiral and twisted chiral vector multiplets in its structure.  Then setting the field strengths to zero in the way described above is just how one enlarges the types of constraints that can be imposed on matter representations.  The observed duality between the chiral and twisted chiral superfields can be seen as the origin of the mirror nature between chiral and twisted chiral vector multiplets described in \cite{gates}.  This observation can made at a more fundamental superspace geometrical level which is described in the next section.
\section{The Discrete Superspace Transformation and Mirror Copies of Supermultiplets}
It is perhaps useful to review the situation with 2d $\mcal{N}=(2,2)$ supersymmetrical representations.
Although it is not widely recognized, the existence of `mirror copies' of 2d $\mcal{N}=(2,2)$ supermultiplets
owes to a very simple supergeometrical circumstance.  We may denote the coordinates of the superspace
by
\be
\label{one}
Z^{M} ~=~ {\big (} \, \theta^{\alpha}; \,  {\bar \theta}{}^{\alpha};   \, \sigma^{m}\, {\big )}  
~=~ {\big (} \, \theta^+, \,  \theta^- ; \,  {\theta}^{\pd} , \,  {\theta}^{\md} ; \, x^{\pp}, \,
x^{\mm} \, {\big )} ~~~~
\ee
and note that it is possible to introduce a discrete transformation (whose generator may be denoted 
by ${\cal M}_m $ [19]) that acts on the coordinates of 2d $\mcal{N}=(2,2)$ superspace
according to
\be
\label{two}
{\cal M}_m :  {\big (} \, \theta^+, \,  \theta^- ; \,  {\theta}^{\pd} , \,  {\theta}^{\md} ; \, x^{\pp}, \,
x^{\mm} \, {\big )} 
 ~=~  {\big (} \, \theta^+, \,  {\theta}^{\md}  ; \,  {\theta}^{\pd} , \,   \theta^- ; \, x^{\pp}, \,
x^{\mm} \, {\big )}.
\ee
If we call the spinor coordinates with $+$ superscripts `right-handed' and those with 
$-$ superscripts `left-handed,' then the effect of this discrete transformation is to act as
the identity on the `right-handed' spinor coordinates as well as bosonic coordinates.  However, it
acts as an outer automorphism acting on the `left-handed' spinor coordinates.

Now such a discrete transformation may or may {\it {not}} constitute a symmetry of any 
particular dynamical $\mcal{N}=(2,2)$ supersymmetric system depending upon what is 
the action upon which it might be applied.  More interesting, however, is that this may not
even be a symmetry of the irreducible representations of $\mcal{N}=(2,2)$ superspace.  In
fact, there are no known irreducible $\mcal{N}=(2,2)$ supermultiplets that realizes this as
a symmetry!  Typically, what happens is that if one begins with a given $\mcal{N}=(2,2)$ 
irreducible representation and applies this discrete symmetry, the result is another distinct 
$\mcal{N}=(2,2)$ irreducible representation, its `mirror image.'

One example of this is provided by considering the 2d $\mcal{N}=(2,2)$ superspace Maxwell
multiplet.   For this purpose we introduce a superspace 
covariant derivative $\nabla_A \equiv D_A ~+~ i g \G_A t$ (in this expression 
$t$ denotes a U(1) Lie algebra generator with an associated commutator algebra
$$ 
[\nabla {}_{\a} , \nabla {}_{\b} \}  ~=~ 0
~~~, ~~~
[\nabla {}_{\a} , \bar \nabla {}_{\b} \} ~=~  i 2 (\g \sp{c}) {}_{\a \b} \nabla
{}_{c} ~+~ ~ 2 g [~ C_{\a \b} S ~-~ i (\g^3 )_{\a \b} P ~] t ~~~,
$$
\begin{equation}  {~~~~~~~~}
\label{three}
[\nabla {}_{\a} , \nabla {}_{b} \} ~=~ g (\g_b )_{\a}{}^{\b} {\bar W}_{\b} t
~~~, ~~~~~
[\nabla {}_{a} , \nabla {}_{b} \} ~=~   - i g \e_{a b} {\cal W} t ~~~, ~~~ 
\end{equation}
and whose consistent Bianchi Identities imply
$$ \nabla_{\a} S ~=~ - i{\bar W}_{\a} ~~~,~~~ \nabla_{\a} P ~=~ -
(\g^3)_{\a}{}^{\b}  {\bar W}_{\b} ~~~,~~~ \nabla_{\a} {\bar W}_{\b}
~=~ 0  ~~~,  ~~~\nabla_{\a} {\rm d} ~=~ (\g^c )_{\a}{}^{\b}  \nabla {}_{c}  
{\bar W}_{\b} ~~~, $$
\begin{equation}
{\nabla}_{\a}  W_{\b} ~=~ i  C_{\a \b} {\rm d} ~-~ 
(\g^3)_{\a \b} {\cal W} ~+~ (\g^a)_{\a \b} ( \nabla_a S)
~-~ i (\g^3 \g^a)_{\a \b} ( \nabla_a P) ~~~.
\end{equation}

Just as the transformation in (\ref{two}) was defined to act on the coordinates of the superspace, an
analogous definition can be realized on the superspace covariant derivative.  For this purpose
it is convenient to first go to a chiral basis defined by
$$
\nabla_+ ~\equiv~ \frac 12 (~1 ~+~ \g^3 )_{\a} {}^{\b} \nabla_{\b} ~~~,~~~ 
\nabla_- ~\equiv~ \frac 12 (~1 ~-~ \g^3 )_{\a} {}^{\b} \nabla_{\b} ~~~,~~~
$$
\begin{equation}
{\bar \nabla}_{\pd} ~\equiv~ \frac 12 (~1 ~+~ \g^3 )_{\a} {}^{\b} {\bar 
\nabla}_{\b} ~~~,~~~ 
{\bar \nabla}_{\md} ~\equiv~ \frac 12 (~1 ~-~ \g^3 )_{\a} {}^{\b} {\bar 
\nabla}_{\b}  ~~~.
\end{equation}
In this chiral basis, the form of the commutator algebra becomes
$$
 \{ \nabla_+ ~,~ \nabla_+ \} ~=~ 0 ~~~,~~~ \{ \nabla_- ~,~ \nabla_- \} ~=~ 0
{}   $$
$$ \{ \nabla_+ ~,~ { \nabla}_- \} ~=~ 0
{}~~~, ~~~ \{ \nabla_+ ~,~ {\nabla_{\md}} \} ~=~ i  {\bar {\P}} \, t
{} $$
$$ \{ \nabla_\pd ~,~ { \nabla}_\md \} ~=~ 0
{}~~~, ~~~ \{ \nabla_\pd ~,~ {\nabla_{-}} \} ~=~ -i   {\P} \, t
{} $$
$$ \{ \nabla_+ ~,~ {\nabla}_{\pd} \} ~=~ i  \nabla_{++} ~~~,~~~
\{ \nabla_- ~,~ {\nabla}_{\md} \} ~=~ i  \nabla_{--}  $$
$$ [ \nabla_+ ~,~ \nabla_{++} ] ~=~ 0 ~~~,~~~ [ \nabla_- ~,~ \nabla_{--} ]
 ~=~ 0    $$
$$ [ \nabla_+ ~,~ \nabla_{--} ] ~=~ -\,   ({\nabla}_- {\bar {\P}} ) \, t
{}  $$
$$ [ \nabla_- ~,~ \nabla_{++} ] ~=~ +   ({\nabla}_+ { {\P}} ) \, t
{}{} $$
\be [ \nabla_{++} ~,~ \nabla_{--} ] ~=~ i  {\cal W} \, t 
 ~~~,  {~~~~}
\ee

Now the superspace covariant derivative $\nabla_A$ is defined just as in (\ref{one})
\be
\label{seven}
\nabla{}_{A} ~=~ {\big (} \, \nabla{}_{\alpha}; \,  {\bar
\nabla}{}{}_{\alpha};   \, \nabla{}_{m}\, {\big )}   ~=~ {\big (} \,
\nabla{}_+, \,  \nabla{}_- ; \,  {\nabla}{}_{\pd} , \,  {\nabla}{}_{\md}
; \, \nabla{}_{++}, \,
\nabla{}_{--} \, {\big )} ~~~~
\ee
and in this form the discrete coordinate transformation may be applied.  This changes the
commutator algebra to the form

$$
 \{ \nabla_+ ~,~ \nabla_+ \} ~=~ 0 ~~~,~~~ \{ \nabla_- ~,~ \nabla_- \} ~=~ 0
{}   $$
$$ \{ \nabla_+ ~,~ { \nabla}_- \} ~=~ - i  {\bar {T}} \, t
{}~~~, ~~~ \{ \nabla_+ ~,~ {\nabla_{\md}} \} ~=~ 0
{} $$
$$ \{ \nabla_\pd ~,~ { \nabla}_\md \} ~=~ + i   {T} \, t
{}~~~, ~~~ \{ \nabla_\pd ~,~ {\nabla_{-}} \} ~=~ 0
{} $$
$$ \{ \nabla_+ ~,~ {\nabla}_{\pd} \} ~=~ i  \nabla_{++} ~~~,~~~
\{ \nabla_- ~,~ {\nabla}_{\md} \} ~=~ i  \nabla_{--}  $$
$$ [ \nabla_+ ~,~ \nabla_{++} ] ~=~ 0 ~~~,~~~ [ \nabla_- ~,~ \nabla_{--} ]
 ~=~ 0    $$
$$ [ \nabla_+ ~,~ \nabla_{--} ] ~=~ 
{}~+~   ({\nabla}_{\md} {\bar {T}} ) \, t  $$
$$ [ \nabla_- ~,~ \nabla_{++} ] ~=~ 
{}~+~   ({\nabla}_{\pd} {\bar {T}} ) \, t
{} $$
\be [ \nabla_{++} ~,~ \nabla_{--} ] ~=~ - i g\,   {\cal W} \, t 
 ~~~,  {~~~~}
\ee
which (after going back to a covariant basis) reads
$$ 
[\nabla {}_{\a} , \nabla {}_{\b} \}  ~=~ i 4 g (\g^3 )_{\a \b} {\bar {T}}
t ~~~, ~~~
[\nabla {}_{\a} , \bar \nabla {}_{\b} \} ~=~  i 2 (\g \sp{c}) {}_{\a \b} \nabla
{}_{c}  ~~~,  ~~~
$$
\begin{equation}
\label{nine}
[\nabla {}_{\a} , \nabla {}_{b} \} ~=~ - g (\g^3 \g_b )_{\a}{}^{\b} {\bar 
\O}_{\b} t ~~~, ~~~
[\nabla {}_{a} , \nabla {}_{b} \} ~=~ 
  - i g \e_{a b} { {\cal W}} t ~~~, ~~~
\end{equation}
and
$$ { ~~~~~~~~ } \nabla_{\a} {\bar {T}} ~=~ 0 ~~~,~~~ {\bar \nabla}_{\a} 
{\bar {T}} ~=~ {\bar \O}_{\b}
  ~~~,~~~ \nabla_{\a} {\bar \O}_{\b}
~=~  i 2 (\g^a)_{\a \b} ( \nabla_a {\bar {T}})  ~~~, $$
\begin{equation} 
{\nabla}_{\a} { \O}_{\b} ~=~  C_{\a \b} ~[ ~
 { {\cal W}} ~+~ i { {\rm d}} ~] ~~~,~~~ \nabla_{\a} { 
{\rm d}} ~=~ (\g^c )_{\a}{}^{\b}  \nabla {}_{c}  {\bar \O}_{\b} ~~~.
\end{equation}
Thus the covariant algebra (\ref{nine}) is the mirror dual of the of the covariant algebra in
(\ref{three}).  What we see is that the semi chiral vector multiplet has a covariant algebra that realizes
a symmetry under the discrete transformation defined in (\ref{seven}).
\section{The Gauge Field Action}
At this point we have only established that the representation is irreducible.  We need to find the action that governs the dynamics for the multiplet.  We can guess the form of the action on dimensional grounds.  Since $[d^4\q]=2$, then the action must be a function of dimensionless fields.  Since the action must also be gauge invariant, this suggest that we can use the mass dimension zero field strengths from the algebra $S$, $P$, and $T$.  What is particularly nice is that we will see a mechanism the theory uses to demonstrate that it is not a direct sum of the chiral and twisted chiral vector multiplets.  
Consider the following actions
\be
S_1=-\frac{1}{4}\int d^2x d^4\q ~S^2
\ee
and
\be
S_2=\frac{1}{2}\int d^2x d^4\q~\bar T T
\ee
Both actions are manifestly supersymmetric since they are written directly in superspace.  However both terms are necessary in order to obtain the field strength squared term $\mcal{W}^2$ in the action.  Lets see how that works.  Evaluating the Grassmann measure with
\be
\int d^4\q=\frac{1}{8}[\nabla^\a\nabla_\a\bar\nabla^\b\bar\nabla_\b+\bar\nabla^\b\bar\nabla_\b\nabla^\a\nabla_\a],
\ee
we get the component actions
\be
S_1=\frac{1}{2}\int d^2x[2i(\bar \l^\b)(\g^a)_\b{}^\a\nabla_a (\l_\a)+\nabla^aS\nabla_aS+\nabla^aP\nabla_aP+(\s_1+\mcal{W})^2+d^2]
\ee
where $W_\a|$=$\l_\a$ and
\be
S_2=\frac{1}{2}\int d^2x[\bar\s \s+2i\bar\r^\b(\g^a)_\b{}^\a\nabla_a \r_\a+4\nabla^a\bar T\nabla_a T]
\ee
with $\O_\a|=\r_\a$ and recall $\s_1=Re(\s)$.  If we just used $S_1$, we would see the e.o.m for $\s_1$ would eliminate the presence of $\mcal{W}$ in the action and thus the gauge field wouldn't have kinetic terms.  We would get the same result if we just used $S_2$ for the more simple reason that $\mcal{W}$ doesn't appear in the action.  It is only the sum of the two terms, $S_1+c_0S_2$, that will generate kinetic terms for the gauge field.  Since each action is separately supersymmetric, the remaining issue to settle is what should the relative coefficient be.  Looking at the kinetic terms for the scalars we see that $c_0$ must be positive with no extra restriction from requiring the appropriate sign for the gauge field kinetic terms in this case $+\mcal{W}^2$.  For simplicity we set $c_0=1$ and consider the action
\bea
S&=&\int d^2x d^4\q[-\frac{1}{4}S^2+\frac{1}{2}\bar T T].\cr
&=&\frac{1}{2}\int d^2x[2i(\bar \l^\b)(\g^a)_\b{}^\a\nabla_a (\l_\a)+\nabla^aS\nabla_aS+\nabla^aP\nabla_aP+(\s_1+\mcal{W})^2+d^2\cr 
\cr
&~&~~~~~~~~~~+\bar\s \s+2i\bar\r^\b(\g^a)_\b{}^\a\nabla_a \r_\a+4\nabla^a\bar T\nabla_a T]
\eea
\subsection{Fayet-Iliopoulos terms}
Since the semi chiral vector multiplet has three real prepotentials, it has room for three F.I. terms \cite{fi} in the action.  They are given by
\bea
S_{FI}&=&\int d^2x[aD^2T+\bar a\bar D^2\bar T+\bar D^\a D_\a(b\P+\bar b\bar\P)]\cr
&=&4\int d^2x[r_1(\s_1+\frac{1}{2}\mcal{W})+r_2\s_2+r_3d].
\eea
The relations between the complex constants $a,b$ and the real constants $r_1,r_2,r_3$ are
\bea
r_1&=&\frac{1}{4}(a+\bar a)+\frac{i}{2}(\bar b-b)\cr
\cr
r_2&=&\frac{i}{4}(a-\bar a)\cr
\cr
r_3&=&\frac{1}{2}(b+\bar b)
\eea

\section{Coupling to Matter}
Coupling the new multiplet to matter can be described in two ways.  One is to evaluate the measure in terms of the covariant derivatives, push the derivatives onto the Kahler potential, and evaluate the fermionic derivatives acting on the matter superfields in terms of the covariantly defined components of the matter superfields.  The other way is to use the prepotentials to adjust the local gauge transformations of the matter field to make the action invariant under the local transformations.  Here we describe the second method because of its greater ability to describe the gauging of target space isometries in non linear sigma models.  Lets recall the process for gauging chiral matter described in \cite{hklr}.  A chiral superfield transforms under the global transformation as
\be
\F\ra e^{i\e t}\F, ~~\bar\F\ra e^{i\e t}\bar\F
\ee
where $\e$ is the constant real transformation parameter.  The kinetic terms for the chiral fields are given by the Kahler potential, $K=K(\bar\F,\F)$ which is invariant under the above transformations.  When the transformation is made local, the parameter $\e$ is promoted to a chiral superfield and thus $\bar\e$ is anti chiral.  However, this means that $\bar\F$ no longer transforms with the same parameter as $\F$ and the invariance of the Kahler potential is lost.  To restore the invariance we need to find a way to get $\bar\F$ and $\F$ to transform with the same transformation parameter.  To do so we use the real prepotential\footnote{This is actually the imaginary part of the complex prepotential for the chiral vector multiplet.}, $V$, from the chiral vector multiplet which transforms as $\d V=i(\bar\e-\e)$.  We define a new field
\be
\tilde\F=e^{-Vt}\bar\F,
\ee
and replace $\bar\F$ in the Kahler potential with $\tilde\F$, i.e. $K=K(\tilde\F,\F)$ and we find that the potential is invariant under local transformations.

A similar procedure will be used to gauge the Kahler potential with left and right semi chiral superfields, however we need to make a few adjustments.  The Kahler potential is a function of the left and right semi chiral superfields, $K=K(\bar X, X, \bar Y, Y)$.  It is invariant under the following transformations.
\be
X\ra e^{i\e t}X, ~~\bar X\ra e^{i\e t}\bar X, ~~Y\ra e^{i\e t}Y, ~~\bar Y\ra e^{i\e t}\bar Y,
\ee
where once again $\e$ is a constant real parameter.  To make the transformation local we, as before, would look to promote $\e$ a superfield.  The issue is choosing the representation to use.  The only consistent choice is to promote the parameter for each superfield to a parameter of the same representation.  The transformations take the form
\be
X\ra e^{-i\L t}X, ~~\bar X\ra e^{-i\bar\L t}\bar X, ~~Y\ra e^{-iUt}Y, ~~\bar Y\ra e^{-i\bar Ut}\bar Y.
\ee
Once again the invariance of the Kahler potential is lost with the above local transformations.  In order to restore the invariance we define new fields using the prepotentials, as before, that will transform properly to restore the invariance of the Kahler potential.  This will happen in a way that looks different from chiral case, though in truth it is actually equivalent.  We recall that the prepotentials actually have two gauge transformations.  We can use the left and right semi chiral transformation of the prepotentials to compensate for the local transformations and exchange them for $L$ gauge transformations\footnote{The procedure described above is a short cut to the procedure we will describe for the case where the fields participating in the gauge transformation are in the same representation i.e. chiral and anti chiral only or some other field and its conjugate.  For a review of this procedure see \cite{superspace}.}.  We define new fields with the prepotentials transforming as in (\ref{prepot1}) and (\ref{prepot2})
\bea
\tilde X&=&e^{iV_1t}X\cr
\cr
\tilde{\bar X}&=&e^{i\bar V_1t}\bar X\cr
\cr
\tilde Y&=&e^{iV_2t}Y\cr
\cr
\tilde{\bar Y}&=&e^{i\bar V_2t}\bar Y
\eea
The new fields all transform with the same parameter and the invariance of the action is restored with the replacements
\be
K(\bar X, X, \bar Y, Y)\ra K(\tilde{\bar X}, \tilde X, \tilde{\bar Y}, \tilde Y).
\ee
The discussion of the gauged action via use of the prepotentials is completed by giving the gauge fixing conditions for the $L$ gauge freedom and choosing the appropriate Wess Zumino gauge.  To start we need to give the components for the left and right semi chiral transformation parameters.
\bea
\label{precomp}
&&\L|=\l,~~~~~~~~~~~~~~~~~~~~~~ U|=u\cr
&&D_\a\L|=\j_\a, ~~~~~~~~~~~~~~~~D_\a U|=\c_\a\cr
&&\bar D_-\L|=\x_-,~~~~~~~~~~~~~~~~ \bar D_- U|=0\cr
&&\bar D_+\L|=0,~~~~~~~~~~~~~~~~~~ \bar D_+ U|=\h_+\cr
&&D^2\L|=F,~~~~~~~~~~~~~~~~~~D^2U|=G\cr
&&\bar D^2\L|=0,~~~~~~~~~~~~~~~~~~~\bar D^2U|=0\cr
&&[D_+,\bar D_+]\L=-i\del_{\pp}\l,~~~~[D_+,\bar D_+]U|=B_{\pp}\cr
&&[D_-,\bar D_-]\L|=C_\mm,~~~~~~~~ [D_-,\bar D_-]U|=-i\del_\mm u\cr
&&[D_-,\bar D_+]\L|=0,~~~~~~~~~~[D_-,\bar D_+]U|=\q'\cr
&&[D_+,\bar D_-]\L|=\q,~~~~~~~~~~[D_+,\bar D_-]U|=0\cr
&&D^2\bar D_+\L|=0,~~~~~~~~~~~~~D^2\bar D_+U|=\o_+\cr
&&D^2\bar D_-\L|=\t_-,~~~~~~~~~~~D^2\bar D_-U|=0\cr
&&\bar D^2D_+\L|=\del_{\pp}\x_-,~~~~~~~\bar D^2D_+U|=0\cr
&&\bar D^2D_-\L|=0,~~~~~~~~~\bar D^2D_-U|=\del_\mm \h_+
\eea
To perform the $L$ gauge fixing we need to decompose the prepotentials into the linear combination of fields that transforms under the $L$ gauge symmetry, and the orthogonal combinations that are inert under the $L$ gauge symmetry.  The combination that $L$ gauge transforms is
\be
\hat V=Re(V_1)+Re(V_2)
\ee
And the orthogonal combinations are
\bea
\tilde V&=&Re(V_2)-Re(V_1)\cr
\tilde V_1&=&Im(V_1)\cr
\tilde V_2&=&Im(V_2)
\eea
We use the $L$ gauge to fix $\hat V=0$.  Then we consider the transformations of the remaining prepotentials components under the remaining gauge transformations to see which we can set to zero in the Wess Zumino gauge.  We set to zero all of the fields that transform by a shift and here give the remaining components.  The gauge field sits in $\tilde V_1$ and $\tilde V_2$ as 
\bea
A_{\pp}&=& -\frac{1}{4}(\g_{\pp})^{++}[D_+,\bar D_+]\tilde V_1|\cr
\cr
A_\mm&=& -\frac{1}{4}(\g_\mm)^{--}[D_-,\bar D_-]\tilde V_2|.
\eea
The remaining components that cannot be set to zero in the Wess Zumino gauge are related to the field strengths given in the algebra (\ref{algebra}) and are given by
\bea
\label{fs}
\frac{i}{4}\bar D^2(\tilde V_2-\tilde V_1)|&=&T|=T\cr
\cr
\frac{i}{2}D_+\bar D_-(\tilde V_2+\tilde V_1)|&=&S-iP|=\P\cr
\cr
\frac{i}{4}\bar D^2D_\a( \tilde V_1+\tilde V_2)|&=&-W_\a|=-\l_\a\cr
\cr
\frac{i}{4}D_\a\bar D^2 \tilde V_2|&=&\O_\a|=\r_\a\cr
\cr
\frac{1}{8}\{D^2,\bar D^2\}\tilde V|&=&(\s_1+\frac{1}{2}\mcal{W})|=(\s_1+\frac{1}{2}\mcal{W})\cr
\cr
\frac{1}{8}\{D^2,\bar D^2\} (\tilde V_2-\tilde V_1)|&=&\s_2|=\s_2\cr
\cr
\frac{1}{8}\{D^2,\bar D^2\} (\tilde V_2+\tilde V_1)|&=&d|=d
\eea
For simplicity we have used the same symbol for the superfield and its lowest component.  This completes the description of the Wess Zumino gauge.
\section{Summary and Conclusion}
In this paper we have observed an new irreducible representation of $\mcal{N}=(2,2)$ supersymmetry, the semi chiral vector multiplet.  This multiplet has the property that its prepotentials transform by semi chiral superfields which was the goal of our original motivation for looking at new vector multiplets.  We have seen that a nice interpretation of semi chiral vector multiplet is that it is the parent multiplet that gives rise to either the chiral or twisted chiral vector multiplet in the appropriate limit.  This is understood from the super geometrical perspective as the realization of a supermultiplet that possess a symmetry under a discrete transformation on the superspace coordinates.  This multiplet should allow for further investigations of the formulation of T duality and the construction of quotients for sigma models with semi chiral superfields \cite{mpv2}.

After the completion of this work, we became aware of similar results obtained by the work of U. Lindstr\"om, M. Ro\v cek, I. Ryb, R. von Unge, and M. Zabzine.  

{\Large\bf Acknowledgments}

We acknowledge support from the UMCP/CSPT for support and W.M. would also like to thank MCTP for hospitality during the final days of this works completion and L. Pando Zayas and D. Vaman for helpful discussions.


\begin{thebibliography}{99}

\bibitem{zumino}
  B.~Zumino,
 ``Supersymmetry And Kahler Manifolds,''
  Phys.\ Lett.\ B {\bf 87} (1979) 203.

\bibitem{ghr}
  S.~J.~Gates, Jr., C.~M.~Hull and M.~Rocek,
 ``Twisted Multiplets And New Supersymmetric Nonlinear Sigma Models,''
  Nucl.\ Phys.\ B {\bf 248} (1984) 157.

\bibitem{gates2}
  S.~J.~Gates, Jr.,
  ``Superspace Formulation Of New Nonlinear Sigma Models,''
  Nucl.\ Phys.\  B {\bf 238}, 349 (1984).

\bibitem{hit}
  N.~Hitchin,
  ``Generalized Calabi-Yau manifolds,''
  Quart.\ J.\ Math.\ Oxford Ser.\  {\bf 54}, 281 (2003)
  [arXiv:math.dg/0209099].
  
 \bibitem{thesis}
M. Gualtieri, "Generalized Complex Geometry", math.DG/0401221.
  
  \bibitem{blr}
  T.~Buscher, U.~Lindstrom and M.~Rocek,
 ``New Supersymmetric Sigma Models With Wess-Zumino Terms,''
  Phys.\ Lett.\ B {\bf 202} (1988) 94.

  \bibitem{lruz}
  U.~Lindstrom, M.~Rocek, R.~von Unge and M.~Zabzine,
``Generalized Kaehler geometry and manifest N = (2,2) supersymmetric
nonlinear sigma-models,''
  JHEP {\bf 0507} (2005) 067
  [arXiv:hep-th/0411186].

\bibitem{lruz2}
  U.~Lindstrom, M.~Rocek, R.~von Unge and M.~Zabzine,
  ``Generalized Kaehler manifolds and off-shell supersymmetry,''
  arXiv:hep-th/0512164.

 \bibitem{hs}
  C.~M.~Hull and B.~J.~Spence,
  ``The gauged nonlinear sigma model with Wess-Zumino term,''
  Phys.\ Lett.\ B {\bf 232}, 204 (1989).

\bibitem{bw}
  J.~Bagger and E.~Witten,
  ``The Gauge Invariant Supersymmetric Nonlinear Sigma Model,''
  Phys.\ Lett.\ B {\bf 118}, 103 (1982).

\bibitem{hklr}
  C.~M.~Hull, A.~Karlhede, U.~Lindstrom and M.~Rocek,
  ``Nonlinear Sigma Models And Their Gauging In And Out Of Superspace,''
  Nucl.\ Phys.\ B {\bf 266}, 1 (1986).

\bibitem{hps}
  C.~M.~Hull, G.~Papadopoulos and B.~J.~Spence,
 ``Gauge symmetries for (p,q) supersymmetric sigma models,''
  Nucl.\ Phys.\ B {\bf 363}, 593 (1991).

\bibitem{rv}
  M.~Rocek and E.~P.~Verlinde,
``Duality, quotients, and currents,''
  Nucl.\ Phys.\ B {\bf 373}, 630 (1992)
  [arXiv:hep-th/9110053].

\bibitem{syz}
  A.~Strominger, S.~T.~Yau and E.~Zaslow,
  ``Mirror symmetry is T-duality,''
  Nucl.\ Phys.\  B {\bf 479}, 243 (1996)
  [arXiv:hep-th/9606040].


\bibitem{mpv}
  W.~Merrell, L.~A.~P.~Zayas and D.~Vaman,
  ``Gauged (2,2) sigma models and generalized Kaehler geometry,''
  arXiv:hep-th/0610116.
  
\bibitem{kt}  
  A.~Kapustin and A.~Tomasiello,
  ``The general (2,2) gauged sigma model with three-form flux,''
  arXiv:hep-th/0610210.

\bibitem{st}
  A.~Sevrin and J.~Troost,
``Off-shell formulation of N = 2 non-linear sigma-models,''
  Nucl.\ Phys.\ B {\bf 492} (1997) 623
  [arXiv:hep-th/9610102].\\
  A.~Sevrin and J.~Troost,
 ``The geometry of supersymmetric sigma-models,''
  arXiv:hep-th/9610103.


\bibitem{gates}
  S.~J.~Gates, Jr.,
   ``Vector multiplets and the phases of N=2 theories in 2-D: Through the
  looking glass,''
  Phys.\ Lett.\ B {\bf 352}, 43 (1995)
  [arXiv:hep-th/9412222].

\bibitem{fi}  
  P.~Fayet and J.~Iliopoulos,
  ``SPONTANEOUSLY BROKEN SUPERGAUGE SYMMETRIES AND GOLDSTONE SPINORS,''
  Phys.\ Lett.\  B {\bf 51}, 461 (1974).

\bibitem{superspace}
  S.~J.~Gates, M.~T.~Grisaru, M.~Rocek and W.~Siegel,
  ``Superspace, or one thousand and one lessons in supersymmetry,''
  Front.\ Phys.\  {\bf 58}, 1 (1983)
  [arXiv:hep-th/0108200].
  
\bibitem{mpv2}
  W.~Merrell, L.~A.~P.~Zayas and D.~Vaman,
  Work In Progress

\end{thebibliography}
 \end{document}